\documentclass[runningheads]{svmult}
\def\plb#1#2#3{{Phys. Lett. B}
{\textbf {#1}},~#2~(#3)}
\def\apjl#1#2#3{{Astrophys. J. Lett.}
{\textbf {#1}},~#2~(#3)}
\def\apj#1#2#3{{Astrophys. Journal}
{\textbf {#1}},~#2~(#3)}
\def\jetpl#1#2#3{{JETP Lett.}
{\textbf {#1}},~#2~(#3)}
\def\jetpsp#1#2#3{{JETP (Sov. Phys.)}
{\textbf {#1}},~#2~(#3)}

\def\jpa#1#2#3{{J. Phys. A}
{\textbf {#1}},~#2~(#3)}
\def\pr#1#2#3{{Phys. Reports}
{\textbf {#1}},~#2~(#3)}
\def\mnras#1#2#3{{Mon. Not. Roy. Astr. Soc.}
{\textbf {#1}},~#2~(#3)}
\def\n#1#2#3{{Nature}
{\textbf {#1}},~#2~(#3)}
\def\cmp#1#2#3{{Commun. Math. Phys.}
{\textbf {#1}},~#2~(#3)}
\def\prsla#1#2#3{{Proc. Roy. Soc. London A}
{\textbf {#1}},~#2~(#3)}
\def\ptp#1#2#3{{Prog. Theor. Phys.}
{\textbf {#1}},~#2~(#3)}
\def\prd#1#2#3{{Phys. Rev. D}
{\textbf {#1}},~#2~(#3)}
\def\prl#1#2#3{{Phys. Rev. Lett.}
{\textbf {#1}},~#2~(#3)}
\def\ibid#1#2#3{{ibid.}
{\textbf {#1}},~#2~(#3)}
\def\npb#1#2#3{{Nucl. Phys. B}
{\textbf {#1}},~#2~(#3)}
\def\jhep#1#2#3{{JHEP}
{\textbf {#1}},~#2~(#3)}
\def\anj#1#2#3{{Astron. J.}
{\textbf {#1}},~#2~(#3)}
\def\baas#1#2#3{{Bull. Am. Astron. Soc.}
{\textbf {#1}},~#2~(#3)}

\begin{document}
\title*{Introduction to Inflationary Cosmology}
\author{George Lazarides}
\institute{Physics Division, School of Technology,
Aristotle University of Thessaloniki, Thessaloniki
54 124, Greece}

\maketitle

\begin{abstract}
The early universe according to the big bang and
the grand unified theories is discussed. The
shortcomings of big bang are summarized together
with their resolution by inflationary cosmology.
Inflation, the subsequent oscillation and decay
of the inflaton, and the resulting `reheating'
of the universe are studied. The density
perturbations produced by inflation and the
temperature fluctuations of the cosmic background
radiation are sketched. The hybrid inflationary
model is described. Two `natural' extensions of
this model which avoid the disaster encountered
in its standard realization from the
overproduction of monopoles are presented.
\end{abstract}

\section{Introduction}
\label{sec:introduction}

\par
The observed Hubble expansion of the universe
together with the discovery of the cosmic microwave
background radiation (CMBR) had established hot big
bang \cite{wkt} as a viable model of the universe.
The success of nucleosynthesis in reproducing the
observed abundance of light elements and the proof
of the black body character of the CMBR then
imposed hot big bang as the standard cosmological
model. This model combined with grand unified
theories (GUTs) \cite{ggps} provides the framework
for discussing the early universe.

\par
Despite its successes, the standard big bang (SBB)
model had some long-standing shortcomings. One of
them is the horizon problem. The CMBR received now
has been emitted from regions which never
communicated before sending light to us. The
question then arises how come the
temperature of the black body radiation from these
regions is so finely tuned as the results of the
cosmic background explorer (COBE) \cite{cobe} show.
Another issue is the flatness problem. The present
universe appears almost flat. This requires that,
in its early stages, the universe was flat with a
great accuracy. Also, combined with GUTs which
predict the existence of superheavy monopoles
\cite{monopole}, the SBB model leads \cite{preskill}
to a catastrophe due to the overproduction of these
monopoles. Finally, the model has no explanation for
the small density perturbations required for the
structure formation in the universe \cite{structure}
and the generation of the observed \cite{cobe}
temperature fluctuations in the CMBR.

\par
Inflation \cite{guth,book,inflation} offers an
elegant solution to all these problems of the SBB
model. The idea is that, in the early universe, a
real scalar field (the inflaton) was displaced from
its vacuum value. If the potential energy density of
this field happens to be quite flat, the roll-over
of the field towards the vacuum can be very slow for
a period of time. During this period, the energy
density is dominated by the almost constant
potential energy density of the inflaton. As a
consequence, the universe undergoes a period of
quasi-exponential expansion, which can readily solve
the horizon and flatness problems by stretching the
distance over which causal contact is established
and reducing any pre-existing curvature in the
universe. It can also adequately dilute the GUT
monopoles. Moreover, it provides us with the
primordial density perturbations which are needed
for explaining the large scale structure in the
universe \cite{structure} as well as the
temperature fluctuations observed in the CMBR.
Inflation can be easily incorporated in GUTs. It
occurs during the GUT phase transition at which the
GUT gauge symmetry breaks by the vacuum expectation
value (vev) of a Higgs field, which also plays the
role of the inflaton.

\par
After the end of inflation, the inflaton oscillates
about the vacuum. The oscillations are damped
because of the dilution of the field energy
density by the cosmological expansion and the
inflaton decay into `light' matter. The resulting
radiation energy density eventually dominates over
the field energy density and the universe returns
to a normal big bang type evolution. The temperature
at which this occurs is historically called `reheat'
temperature although there is neither supercooling
nor reheating of the universe \cite{reheat}.

\par
An important disadvantage of the early realizations
of inflation is that they require tiny parameters in
order to reproduce the COBE results on the CMBR. To
solve this `naturalness' problem, hybrid inflation
has been introduced \cite{hybrid}. The idea was to
use two real scalar fields instead of one that was
normally used. One field may be a gauge non-singlet
and provides the `vacuum' energy density which
drives inflation, while the other is the slowly
varying field during inflation. This splitting of
roles between two fields allows us to reproduce the
temperature fluctuations of the CMBR with `natural'
(not too small) values of the relevant parameters.
Hybrid inflation, although initially introduced in
the context of non-supersymmetric GUTs, can be
`naturally' incorporated \cite{lyth,dss} in
supersymmetric (SUSY) GUTs.

\par
Unfortunately, the monopole problem reappears in
hybrid inflation. The end of inflation, in this
case, is abrupt and is followed by a `waterfall'
regime during which the system falls towards the
vacuum manifold and performs damped oscillations
about it. If the vacuum manifold is homotopically
non-trivial, topological defects will be
copiously formed \cite{smooth} by the Kibble
mechanism \cite{kibble} since the system can end
up at any point of this manifold with equal
probability. So a disaster is encountered in the
hybrid inflationary models which are based on a
gauge symmetry breaking predicting monopoles.

\par
One idea \cite{smooth,jean,shi,talks} for solving
the monopole problem of SUSY hybrid inflation is
to include into the standard superpotential for
hybrid inflation the leading non-renormalizable
term. This term cannot be excluded by any
symmetries and, if its dimensionless coefficient
is of order unity, can be comparable with the
trilinear coupling of the standard superpotential
(whose coefficient is $\sim 10^{-3}$). Actually,
we have two options. We can either keep
\cite{jean} both these terms or remove
\cite{smooth,shi} the trilinear term by a
discrete symmetry and keep only the leading
non-renormalizable term. The pictures emerging
in the two cases are different. However, they
share a common feature. The GUT gauge group is
broken during inflation and, thus, no topological
defects can form at the end of inflation. So, the
monopole problem is solved.

\section{The Big Bang Model}
\label{sec:bigbang}

\par
We will start with an introduction to the salient
features of the SBB model \cite{wkt} and a
summary of the history of the early universe in
accordance to GUTs.

\subsection{Hubble Expansion}
\label{subsec:hubble}

\par
At cosmic times $t\stackrel{_{>}}{_{\sim }}
t_{P}\equiv M_{P}^{-1}\sim 10^{-44}~{\rm{sec}}$
($M_{P}=1.22\times 10^{19}~{\rm{GeV}}$ is the
Planck scale) after the big bang, the quantum
fluctuations of gravity are suppressed and
classical relativity is adequate. Strong, weak
and electromagnetic interactions, however,
require quantum field theoretic treatment.

\par
We assume that the universe is homogeneous and
isotropic. The strongest evidence for this
{\it cosmological principle} is the observed
\cite{cobe} isotropy of the CMBR. The space-time
metric then takes the Robertson-Walker form
\begin{equation}
ds^{2}=-dt^{2}+ a^{2}(t)\left[\frac{dr^{2}}
{1-kr^2}+r^{2}(d\theta^{2}+\sin^{2}\theta~
d\varphi^{2}) \right],
\label{eq:rw}
\end{equation}
where $r$, $\varphi$ and $\theta$ are `comoving'
polar coordinates, which remain fixed for objects
that just follow the general cosmological expansion.
$k$ is the `scalar curvature' of the 3-space and
$k=0$, $>0$ or $<0$ corresponds to flat, closed or
open universe. The dimensionless parameter $a(t)$ is
the `scale factor' of the universe. We take $a_{0}
\equiv a(t_{0})=1$, where $t_{0}$ is the present
cosmic time.

\par
The `instantaneous' radial physical distance is
given by
\begin{equation}
R=a(t)\int_{0}^{r}\frac{dr}{(1-kr^{2})^\frac{1}{2}}
~\cdot
\label{eq:dist}
\end{equation}
For flat universe ($k=0$), $\bar{R}=a(t)\bar{r}$
($\bar{r}$ is a `comoving' and  $\bar{R}$ a physical
radial vector in 3-space) and the velocity of an
object is
\begin{equation}
\bar{V}=\frac{d\bar{R}}{dt}=\frac{\dot{a}}{a}\bar{R}
+a\frac{d\bar{r}}{dt}~, \label{eq:velocity}
\end{equation}
where overdots denote derivation with respect to $t$.
The second term in the right hand side (rhs) of this
equation is the `peculiar velocity', $\bar{v}= a(t)
\dot{\bar{r}}$, of the object, i.e., its velocity
with respect to the `comoving' coordinate system. For
$\bar{v}=0$, (\ref{eq:velocity}) becomes
\begin{equation}
\bar{V}=\frac{\dot{a}}{a}\bar{R}\equiv H(t)\bar{R}~,
\label{eq:hubblelaw}
\end{equation}
where $H(t)\equiv\dot{a}(t)/a(t)$ is the Hubble
parameter. This is the well-known Hubble law
asserting that all objects run away from each other
with velocities proportional to their distances and
is the first success of SBB cosmology.

\subsection{Friedmann Equation}
\label{subsec:friedmann}

\par
In a homogeneous and isotropic universe, the energy
momentum tensor takes the form
$(T_{\mu}^{~\nu})={\rm{diag}}(-\rho, p, p, p)$,
where $\rho$ is the energy density and $p$ the
pressure. Energy momentum conservation then yields
the continuity equation
\begin{equation}
\frac{d\rho}{dt}=-3H(t)(\rho+p)~,
\label{eq:continuity}
\end{equation}
where the first term in the rhs describes the
dilution of the energy due to the Hubble expansion
and the second term the work done by pressure.

\par
For a universe described by the metric in
(\ref{eq:rw}), Einstein's equations
\begin{equation}
R_{\mu}^{~\nu}-\frac{1}{2}~\delta_{\mu}^{~\nu}R=
8\pi G~T_{\mu}^{~\nu},
\label{eq:einstein}
\end{equation}
where $R_{\mu}^{~\nu}$ and $R$ are the Ricci tensor
and scalar curvature and $G\equiv M_{P}^{-2}$ is the
Newton's constant, lead to the Friedmann equation
\begin{equation}
H^{2}\equiv \left(\frac{\dot{a}(t)}{a(t)}\right)^{2}
=\frac{8\pi G}{3}\rho-\frac{k}{a^{2}}~\cdot
\label{eq:friedmann}
\end{equation}

\par
Averaging $p$, we write $\rho+p=(1+w)\rho\equiv
\gamma\rho$ and (\ref{eq:continuity}) gives $\rho
\propto a^{-3\gamma}$. For a universe dominated
by pressureless matter, $\gamma=1$ and, thus,
$\rho\propto a^{-3}$. This is interpreted as mere
dilution of a fixed number of particles in a
`comoving' volume due to the Hubble expansion. For
a radiation dominated universe, $p=\rho/3$ and,
thus, $\gamma=4/3$, which gives $\rho\propto
a^{-4}$. The extra factor of $a(t)$ is due to the
red-shifting of all wave lengths by the expansion.
Substituting $\rho
\propto a^{-3 \gamma}$ in (\ref{eq:friedmann})
with $k=0$, we get $a(t)\propto t^{2/3\gamma}$
which, for $a(t_{0})=1$, gives
\begin{equation}
a(t)=\left(\frac{t}{t_0}\right)^\frac{2}{3\gamma}.
\label{eq:expan}
\end{equation}
For `matter' or `radiation', we obtain
$a(t)=(t/t_{0})^{2/3}$ or $a(t)=(t/t_{0})^{1/2}$.

\par
The early universe is radiation dominated and its
energy density is
\begin{equation}
\rho=\frac{\pi^{2}}{30}\left(N_{b}+\frac{7}{8}
N_{f}\right)T^{4}\equiv c~T^{4},
\label{eq:boltzman}
\end{equation}
where $T$ is the cosmic temperature and $N_{b(f)}$
the number of massless bosonic (fermionic) degrees
of freedom. The quantity $g_{*}=N_{b}+(7/8)N_{f}$
is called effective number of massless degrees of
freedom. The entropy density is
\begin{equation}
s=\frac{2\pi^{2}}{45}~g_{*}~T^{3}.
\label{eq:entropy}
\end{equation}
Assuming adiabatic universe evolution, i.e.,
constant entropy in a `comoving' volume
($sa^{3}={\rm{constant}}$), we obtain
$aT={\rm{constant}}$. The temperature-time relation
during radiation dominance is then derived from
(\ref{eq:friedmann}) (with $k=0$):
\begin{equation}
T^{2}=\frac{M_{P}}{2(8\pi c/3)^\frac{1}{2}t}~\cdot
\label{eq:temptime}
\end{equation}
Classically, the expansion starts at $t=0$ with
$T=\infty$ and $a=0$. This initial singularity is,
however, not physical since general relativity
fails for $t\stackrel{_{<}}{_{\sim }}t_{P}$ (the
Planck time). The only meaningful statement is that
the universe, after a yet unknown initial stage,
emerges at $t\sim t_{P}$ with $T\sim M_{P}$.

\subsection{Important Cosmological Parameters}
\label{subsec:parameter}

\par
The most important parameters describing the
expanding universe are:
\begin{list}
\setlength{\rightmargin=0cm}{\leftmargin=0cm}
\item[{\bf i.}] The present value of the Hubble
parameter (known as Hubble constant) $H_{0}
\equiv H(t_{0})=100~h~\rm{km}~\rm{sec}^{-1}~
\rm{Mpc}^{-1}$ ($h\approx 0.72\pm 0.07$
\cite{h}).
\vspace{.25cm}
\item[{\bf ii.}] The fraction $\Omega=\rho/
\rho_{c}$, where $\rho_{c}$ is the critical
density corresponding to a flat universe. From
(\ref{eq:friedmann}), $\rho_{c}=3H^{2}/8\pi G$
and $\Omega=1+k/a^{2}H^{2}$. $\Omega=1$, $>1$
or $<1$ corresponds to flat, closed or open
universe. Assuming inflation (see below), the
present value of $\Omega$ must be $\Omega_{0}=1$
in accord with the recent DASI observations
which yield \cite{dasi} $\Omega_{0}=1\pm 0.04$.
The low deuterium abundance measurements
\cite{deuterium} give $\Omega_{B}h^2\approx
0.020\pm 0.001$, where $\Omega_{B}$ is the
baryonic contribution to $\Omega_0$. This result
implies that $\Omega_{B}\approx 0.039\pm 0.077$.
The total contribution $\Omega_M$ of matter to
$\Omega_0$ can then be determined from the
measurements \cite{cluster} of the
baryon-to-matter ratio in clusters. It is found
that $\Omega_M\approx 1/3$, which shows that
most of the matter in the universe is
non-baryonic, i.e., dark matter. Moreover, we
see that about $2/3$ of the energy density of the
universe is not even in the form of matter and we
call it dark energy.
\vspace{.25cm}
\item[{\bf iii.}] The deceleration parameter
\begin{equation}
q=-\frac{(\ddot{a}/\dot{a})}{(\dot{a}/a)}
=\frac{\rho+3p}{2\rho_{c}}~\cdot \label{eq:decel}
\end{equation}
Measurements of type Ia supernovae \cite{lambda}
indicate that the universe is speeding up ($q_0<0$).
This requires that, at present, $p<0$ as can be
seen from (\ref{eq:decel}). Negative pressure can
only be attributed to the dark energy since matter
is pressureless. Equation (\ref{eq:decel}) gives
$q_0=(\Omega_0+3w_X\Omega_X)/2$, where
$\Omega_X=\rho_X/\rho_c$ and $w_X=p_X/\rho_X$ with
$\rho_X$ and $p_X$ being the dark energy density
and pressure. Observations prefer $w_X=-1$, with
a 95\% confidence limit $w_X<-0.6$ \cite{w}. Thus,
dark energy can be interpreted as something close
to a non-zero cosmological constant (see below).
\end{list}

\subsection{Particle Horizon}
\label{subsec:parhor}

\par
Light travels only a finite distance from the time
of big bang ($t=0$) until some cosmic time $t$.
From (\ref{eq:rw}), we find that the propagation
of light along the radial direction is described by
$a(t)dr=dt$. The particle horizon, which is the
`instantaneous' distance at $t$ travelled by light
since $t=0$, is then
\begin{equation}
d_{H}(t)=a(t)\int_{0}^{t}
\frac{dt^{\prime}}{a(t^{\prime})}~\cdot
\label{eq:hor}
\end{equation}
The particle horizon is an important notion since
it coincides with the size of the universe already
seen at time $t$ or, equivalently, with the distance
at which causal contact has been established at $t$.
Equations (\ref{eq:expan}) and (\ref{eq:hor}) give
\begin{equation}
d_{H}(t)=\frac{3\gamma}{3\gamma-2}t~,
~\gamma\neq \frac{2}{3}~.
\label{eq:hort}
\end{equation}
Also,
\begin{equation}
H(t)=\frac{2}{3\gamma}t^{-1},
~d_{H}(t)=\frac{2}{3\gamma-2}H^{-1}(t)~.
\label{eq:hubblet}
\end{equation}
For `matter' (`radiation'), these formulae become
$d_{H}(t)=2H^{-1}(t)=3t$ ($d_{H}(t)=H^{-1}(t)=2t$).
Assuming matter dominance, the present particle
horizon (cosmic time) is $d_{H}(t_{0})=2H_{0}^{-1}
\approx 6,000~h^{-1}~{\rm{Mpc}}$ ($t_{0}=
2H_{0}^{-1}/3\approx 6.5\times 10^{9} ~h^{-1}~
{\rm{years}}$). The present $\rho_{c}=3H_{0}^{2}/
8\pi G\approx 1.9\times 10^{-29}~h^2
~{\rm{gm/cm^{3}}}$.

\subsection{Brief History of the Early Universe}
\label{subsec:history}

\par
We will now summarize the early universe evolution
according to GUTs \cite{ggps}. We take a GUT gauge
group $G$ ($=SU(5)$, $SO(10)$, $SU(3)^{3}$, ...)
with or without SUSY. At a scale $M_{X}\sim 10^{16}
~{\rm{GeV}}$ (the GUT mass scale), $G$ breaks to
the standard model gauge group $G_{S}$ by the vev
of an appropriate Higgs field $\phi$. (For
simplicity, we take this breaking occurring in one
step.) $G_{S}$ is, subsequently, broken to
$SU(3)_{c}\times U(1)_{em}$ at the electroweak
scale $M_{W}$.

\par
GUTs together with SBB provide a suitable
framework for discussing the early universe for
$t\stackrel{_{>}}{_{\sim }} 10^{-44}~{\rm{sec}}$.
They predict that the universe, as it expands and
cools, undergoes \cite{kl} a series of phase
transitions during which the gauge symmetry is
gradually reduced and important phenomena occur.

\par
After the big bang, $G$ was unbroken and the
universe was filled with a hot `soup' of massless
particles which included photons, quarks, leptons,
gluons, $W^{\pm}$, $Z^{0}$, the GUT gauge bosons
$X$, $Y$, ... and several Higgs bosons. (In the
SUSY case, the SUSY partners were also present.)
At $t\sim 10^{-37}~{\rm{sec}}$ ($T\sim 10^{16}~
{\rm{GeV}}$), $G$ broke down to $G_{S}$ and the
$X$, $Y$,... and some Higgs bosons acquired
masses $\sim M_{X}$. Their out-of-equilibrium
decay could, in principle, produce
\cite{dimopoulos,bau} the observed baryon
asymmetry of the universe. Important ingredients
are the violation of baryon number, which is
inherent in GUTs, and C and CP violation. This is
the second (potential) success of SBB.

\par
During the GUT phase transition, topologically
stable extended objects \cite{kibble} such as
monopoles \cite{monopole}, cosmic strings
\cite{string} or walls \cite{wall} can also be
produced. Monopoles, which exist in most GUTs,
can lead into problems \cite{preskill} which are,
however, avoided by inflation
\cite{guth,book,inflation} (see
Sects.\ref{subsec:monopole} and \ref{subsec:infmono}).
This is a period of an exponentially fast expansion
of the universe which can occur during some GUT phase
transition. Cosmic strings can contribute \cite{zel}
to the density perturbations needed for structure
formation \cite{structure} in the universe, whereas
walls are \cite{wall} catastrophic and GUTs should be
constructed so that they avoid them (see e.g.,
\cite{axion}) or inflation should extinguish them.

\par
At $t\sim 10^{-10}~{\rm{sec}}$ or $T\sim 100
~{\rm{GeV}}$, the electroweak transition takes place
and $G_{S}$ breaks to $SU(3)_{c}\times U(1)_{em}$.
$W^{\pm}$, $Z^{0}$ and the electroweak Higgs fields
acquire masses $\sim M_{W}$. Subsequently, at $t\sim
10^{-4}~{\rm{sec}}$ or $T\sim 1~{\rm{GeV}}$, color
confinement sets in and the quarks get bounded forming
hadrons.

\par
The direct involvement of particle physics essentially
ends here since most of the subsequent phenomena fall
into the realm of other branches. We will, however,
sketch some of them since they are crucial for
understanding the earlier stages of the universe
evolution where their origin lies.

\par
At $t\approx 180~{\rm{sec}}$ ($T\approx 1~{\rm{MeV}}$),
nucleosynthesis takes place, i.e., protons and neutrons
form nuclei. The abundance of light elements ($D$,
$^{3}He$, $^{4}He$ and $^{7}Li$) depends \cite{peebles}
crucially on the number of light particles (with mass
$\stackrel{_{<}}{_{\sim }} 1~{\rm{MeV}}$), i.e., the
number of light neutrinos, $N_{\nu}$, and
$\Omega_{B}h^{2}$. Agreement with observations
\cite{deuterium} is achieved for $N_{\nu}=3$ and
$\Omega_{B}h^{2}\approx 0.020$. This is the third
success of SBB cosmology. Much later, at the so-called
`equidensity' point, $t_{\rm{eq}}\approx 5\times
10^4~{\rm{years}}$, matter dominates over radiation.

\par
At $t\approx 200,000~h^{-1} {\rm{years}}$ ($T\approx
3,000~{\rm{K}}$), the `decoupling' of matter and
radiation and the `recombination' of atoms occur.
After this, radiation evolves as an independent
component of the universe and is detected today as
CMBR with temperature $T_{0}\approx 2.73~{\rm{K}}$.
The existence of the CMBR is the fourth success of
SBB. Finally, structure formation \cite{structure}
starts at $t\approx 2\times 10^{8}~{\rm{years}}$.

\section{Shortcomings of Big Bang}
\label{sec:short}

\par
The SBB model has been successful in explaining,
among other things, the Hubble expansion, the
existence of the CMBR and the abundance of the
light elements formed during nucleosynthesis.
Despite its successes, this model had a number
of long-standing shortcomings which we will now
summarize:

\subsection{Horizon Problem}
\label{subsec:horizon}

\par
The CMBR, which we receive now, was emitted at
the time of `decoupling' of matter and radiation
when the cosmic temperature was $T_d\approx 3,000
~\rm{K}$. The decoupling time, $t_d$, can be
calculated from
\begin{equation}
\frac{T_0}{T_d}=\frac{2.73~\rm{K}}{3,000~\rm{K}}
=\frac{a(t_d)}{a(t_0)}=\left(\frac {t_d}{t_0}
\right)^\frac{2}{3}\cdot
\label{eq:dec}
\end{equation}
It turns out that $t_d \approx 200,000~h^{-1}$
years.

\par
The distance over which the CMBR has travelled
since its emission is
\begin{equation}
a(t_0)\int^{t_{0}}_{t_{d}}\frac{dt^\prime}
{a(t^\prime)}=3t_0\left[1-\left(\frac{t_d}{t_0}
\right)^\frac{2}{3}\right]\approx 3t_0\approx
6,000~h^{-1}~\rm{Mpc}~,
\label{eq:lss}
\end{equation}
which coincides with $d_H(t_0)$. A sphere
of radius $d_H(t_0)$ around us is called
the `last scattering surface' since the CMBR
has been emitted from it. The particle horizon
at $t_d$, $3t_d\approx 0.168~h^{-1}~\rm{Mpc}$,
expanded until now to become $0.168~h^{-1}
(a(t_0)/a(t_d))~{\rm{Mpc}}\approx 184~h^{-1}
~\rm{Mpc}$. The angle subtended by this
`decoupling' horizon now is $\theta_{d}\approx
184/6,000\approx 0.03~\rm{rads}$. Thus, the
sky splits into $4\pi/(0.03)^2\approx 14,000$
patches which never communicated before
emitting the CMBR. The puzzle then is how can
the temperature of the black body radiation
from these patches be so finely tuned as COBE
\cite{cobe} requires.

\subsection{Flatness Problem}
\label{subsec:flatness}

\par
The present energy density of the universe has
been observed \cite{dasi} to be very close to
its critical value corresponding to a flat
universe ($\Omega_{0}=1\pm 0.04$). From
(\ref{eq:friedmann}), we obtain $(\rho-\rho_c)
/\rho_c=3(8\pi G\rho_c)^{-1}(k/a^2)\propto a$
for `matter'. Thus, in the early universe,
$|(\rho-\rho_c)/\rho_c|\ll 1$ and the question
is why the initial energy density of the
universe was so finely tuned to its critical
value.

\subsection{Magnetic Monopole Problem}
\label {subsec:monopole}

\par
This problem arises only if we combine SBB with
GUTs \cite{ggps} which predict the existence of
monopoles. According to GUTs, the universe
underwent \cite{kl} a (second order) phase
transition during which an appropriate Higgs
field, $\phi$, developed a non-zero vev and the
GUT gauge group, $G$, broke to $G_{S}$.

\par
The GUT phase transition produces monopoles
\cite{monopole}. They are localized deviations
from the vacuum with radius $\sim M_X^{-1}$ and
mass $m_M\sim M_X/\alpha_G$ ($\alpha_G=
g^2_{G} /4\pi$, where $g_G$ is the GUT gauge
coupling constant). The value of $\phi$ on a
sphere, $S^2$, of radius $\gg M_{X}^{-1}$ around
the monopole lies on the vacuum manifold $G/G_S$
and we, thus, obtain a mapping:
$S^2\rightarrow G/G_S$. If this mapping is
homotopically non-trivial, the monopole is
topologically stable.

\par
The initial `relative' monopole number density
must satisfy the causality bound \cite{einhorn}
$r_{M,\rm{in}}=(n_M/T^3)_{\rm{in}}
\stackrel{_{>}}{_{\sim }} \rm{10^{-10}}$, which
comes from the requirement that, at
monopole production, $\phi$ cannot be
correlated at distances bigger than the particle
horizon. The subsequent evolution
of monopoles is studied in \cite{preskill}. The
result is that, if $r_{M,\rm{in}}
\stackrel{_{>}}{_{\sim }}\rm{10^{-9}}$
($\stackrel{_{<}}{_{\sim }}\rm{10^{-9}}$), the
final `relative' monopole number density
$r_{M,\rm{fin}}\sim 10^{-9}$
($\sim r_{M,\rm{in}}$). This combined with the
causality bound yields $r_{M,\rm{fin}}
\stackrel{_{>}}{_{\sim }} \rm{10^{-10}}$. However,
the requirement that monopoles do not dominate
the energy density of the universe at
nucleosynthesis gives
\begin{equation}
r_M (T \approx 1~\rm{MeV})
\stackrel{_{<}} {_{\sim}}\rm{10^{-19}},
\label{eq:nucleo}
\end{equation}
and we obtain a clear discrepancy of about nine
orders of magnitude.

\subsection{Density Perturbations}
\label{subsec:fluct}

\par
For structure formation \cite{structure} in the
universe, we need a primordial density perturbation,
$\delta \rho/\rho$, at all length scales with a
nearly flat spectrum \cite{hz}. We also need an
explanation of the temperature fluctuations of the
CMBR observed by COBE \cite{cobe} at angles $\theta
\stackrel{_{>}}{_{\sim }}\theta_d\approx 2^{o}$
which violate causality (see
Sect.\ref{subsec:horizon}).

\section{Inflation}
\label{sec:inflation}

\par
The above four cosmological puzzles are solved
by inflation \cite{guth,book,inflation}.
Take a real scalar field $\phi$ (the inflaton)
with (symmetric) potential energy density
$V(\phi)$ which is quite flat near $\phi=0$
and has minima at $\phi=\pm\langle\phi\rangle$
with $V(\pm\langle\phi\rangle)=0$. At high
$T$'s, $\phi=0$ due to the temperature
corrections to $V(\phi)$. As $T$ drops, the
effective potential tends to the $T$=0
potential but a small barrier separating the
local minimum at $\phi=0$ and the vacua at
$\phi=\pm\langle\phi\rangle$ remains. At some
point, $\phi$ tunnels out to
$\phi_1\ll\langle\phi\rangle$ and a bubble
with $\phi=\phi_1$ is created. The field then
rolls over to the minimum of $V(\phi)$ very
slowly (due to the flatness of $V(\phi)$) with
the energy density $\rho\approx V(\phi=0)
\equiv V_0$ remaining practically constant for
quite some time. The Lagrangian density
\begin{equation}
L=\frac{1}{2}\partial_{\mu}\phi
\partial^{\mu}\phi -V(\phi)
\label{eq:lagrange}
\end{equation}
gives the energy momentum tensor
\begin{equation}
T_{\mu}^{~\nu}=-\partial_\mu\phi\partial^\nu
\phi+\delta_{\mu}^{~\nu}\left(\frac{1}{2}
\partial_\lambda\phi\partial^\lambda\phi-
V(\phi)\right),
\label{eq:energymom}
\end{equation}
which during the roll-over becomes
$T_{\mu}^{~\nu}\approx -V_{0}~\delta_{\mu}^{~\nu}$
yielding $\rho\approx -p\approx V_0$. So, the
pressure is opposite to the energy density
in accord with (\ref{eq:continuity}). $a(t)$
grows (see below) and the `curvature term',
$k/a^2$, in (\ref{eq:friedmann}) diminishes. We
thus get
\begin{equation}
H^2\equiv\left(\frac{\dot{a}}{a}\right)^2=
\frac{8\pi G}{3}V_0~,
\label{eq:inf}
\end{equation}
which gives $a(t)\propto e^{Ht}$, $H^2=
(8\pi G/3)V_0={\rm constant}$. So the bubble
expands exponentially for some time and $a(t)$
grows by a factor
\begin{equation}
\frac {a(t_f)}{a(t_i)}={\rm{exp}}H(t_f-t_i)
\equiv{\rm{exp}}H\tau~, \label{eq:efold}
\end{equation}
between an initial ($t_i$) and a final ($t_f$)
time.

\par
The scenario just described is known as `new'
\cite{new} inflation. Alternatively, we can imagine,
at $t_{P}$, a region of size $\ell_{P}\sim M_{P}^{-1}$
(the Planck length) where the inflaton is large and
almost uniform carrying negligible kinetic energy.
This region can inflate (exponentially expand) as
$\phi$ rolls down towards the vacuum. This scenario
is called `chaotic' \cite{chaotic} inflation.

\par
We will now show that, with an adequate number of
e-foldings, $N=H\tau$, the first three cosmological
puzzles are easily resolved (we leave the question
of density perturbations for later).

\subsection{Resolution of the Horizon Problem}
\label{subsec:infhor}

\par
The particle horizon during inflation
\begin{equation}
d_H(t)=e^{Ht}\int^t_{t_{i}}
\frac{d t^\prime}{e^{Ht^\prime}}\approx
H^{-1}{\rm{exp}}H(t-t_i)~,
\label{eq:horizon}
\end{equation}
for $t-t_i\gg H^{-1}$, grows as fast as $a(t)$. At
$t_f$, $d_H(t_f)\approx H^{-1}{\rm{exp}}H\tau$ and
$\phi$ starts oscillating about the vacuum. It
then decays and `reheats' \cite{reheat} the
universe at a temperature $T_r\sim 10^9~{\rm{GeV}}$
\cite{gravitino} after which normal big bang
cosmology is recovered. $d_H(t_{f})$ is stretched
during the $\phi$-oscillations by a factor
$\sim 10^9$ and between $T_r$ and now by a
factor $T_r/T_0$. So, it finally becomes
$H^{-1}e^{H\tau}10^9(T_r/T_0)$,
which must exceed $2H_{0}^{-1}$ if the horizon
problem is to be solved. This readily holds for
$V_0\approx M_{X}^{4}$,
$M_{X}\sim 10^{16}~{\rm GeV}$
and $N\stackrel{_{>}}{_{\sim }}55$.

\subsection{Resolution of the Flatness Problem}
\label{subsec:infflat}

\par
The `curvature term' of the Friedmann equation, at
present, is given by
\begin{equation}
\frac{k}{a^2}\approx\left(\frac{k}{a^2}\right)_{bi}
e^{-2H\tau}~10^{-18}\left(\frac {10^{-13}~{\rm{GeV}}}
{10^9~{\rm{GeV}}}\right)^2,
\label{eq:curvature}
\end{equation}
where the terms in the rhs are the `curvature term'
before inflation, and its growth factors during
inflation, $\phi$-oscillations and after `reheating'.
Assuming $(k/a^2)_{bi}\sim H ^2$, we get $\Omega_0-1
=k/a_{0}^{2}H_{0}^{2}\sim 10^{48}~e^{-2H \tau}\ll 1$
for $H\tau\gg 55$. Strong inflation implies that the
present universe is flat with a great accuracy.

\subsection{Resolution of the Monopole Problem}
\label{subsec:infmono}

\par
For $N\stackrel{_{>}}{_{\sim }} 55$, the monopoles
are diluted by at least 70 orders of magnitude and
become irrelevant. Also, since $T_r \ll m_M$, there
is no monopole production after `reheating'.
Extinction of monopoles may also be achieved by
non-inflationary mechanisms such as magnetic
confinement \cite{fate}. For models leading to a
possibly measurable monopole density see e.g.,
\cite{thermal,trinification}.

\section{Detailed Analysis of Inflation}
\label{sec:detail}

The Hubble parameter during inflation depends on
the value of $\phi$:
\begin{equation}
H^{2}(\phi)=\frac{8\pi G}{3}V(\phi)~.
\label{eq:hubble}
\end{equation}
To find the evolution equation for $\phi$ during
inflation, we vary the action
\begin{equation}
\int\sqrt{-{\rm{det}}(g)}~d^{4}x\left(\frac{1}{2}
\partial_ {\mu}\phi\partial^{\mu}\phi-V(\phi)+
M(\phi)\right),
\label{eq:action}
\end{equation}
where $g$ is the metric tensor and $M(\phi)$ the
(trilinear) coupling of $\phi$ to `light' matter
causing its decay. Assuming that this coupling
is weak, one finds \cite{dh}
\begin{equation}
\ddot{\phi}+3H\dot{\phi}+\Gamma_{\phi}\dot{\phi}+
V^{\prime}(\phi)=0~,
\label{eq:evolution}
\end{equation}
where the prime denotes derivation with respect
to $\phi$ and $\Gamma_{\phi}$ is the decay width
\cite{width} of the inflaton. Assume, for the
moment, that the decay time of $\phi$, $t_d=
\Gamma_{\phi}^{-1}$, is much greater than $H^{-1}$,
the expansion time for inflation. Then the term
$\Gamma_{\phi}\dot{\phi}$ can be ignored and
(\ref{eq:evolution}) becomes
\begin{equation}
\ddot{\phi}+3H\dot{\phi}+V^{\prime}(\phi)=0~.
\label{eq:reduce}
\end{equation}
Inflation is by definition the situation where
$\ddot{\phi}$ is subdominant to the `friction term'
$3H\dot{\phi}$ (and the kinetic energy density is
subdominant to the potential one). Equation
(\ref{eq:reduce}) then reduces to the inflationary
equation \cite{slowroll}
\begin{equation}
3H\dot{\phi}=-V^{\prime}(\phi)~,
\label{eq:infeq}
\end{equation}
which gives
\begin{equation}
\ddot{\phi}=-\frac{V^{\prime\prime}(\phi)\dot{\phi}}
{3H(\phi)}+\frac{V^{\prime}(\phi)}
{3H^{2}(\phi)}H^\prime(\phi)\dot{\phi}~.
\label{eq:phidd}
\end{equation}
Comparing the two terms in the rhs of this equation
with the `friction term' in (\ref{eq:reduce}), we
get the conditions for inflation (slow roll
conditions):
\begin{equation}
\epsilon~,~|\eta|\leq 1~,~{\rm with}~\epsilon\equiv
\frac{M_{P}^{2}}{16\pi}\left(\frac{V^{\prime}(\phi)}
{V(\phi)}\right)^{2},~\eta\equiv\frac{M_{P}^{2}}
{8\pi}\frac{V^{\prime\prime}(\phi)}{V(\phi)}~\cdot
\label{eq:src}
\end{equation}
The end of the slow roll-over occurs when either of
these inequalities is saturated. If $\phi_f$ is the
value of $\phi$ at the end of inflation, then $t_f
\sim H^{-1}(\phi_f)$.

\par
The number of e-foldings during inflation can be
calculated as follows:
\begin{equation}
N(\phi_{i}\rightarrow \phi_{f})\equiv\ln
\left(\frac{a(t_{f})}
{a(t_{i})}\right)=\int^{t_{f}} _{t_{i}} Hdt=
\int^{\phi_{f}}_{\phi_{i}}\frac{H (\phi)}
{\dot{\phi}}d\phi=-\int^{\phi_{f}}_{\phi_{i}}
\frac {3 H^2 (\phi) d \phi}{V^{\prime}(\phi)},
\label{eq:nefolds}
\end{equation}
where (\ref{eq:efold}), (\ref{eq:infeq}) were used.
We shift $\phi$ so that the global minimum of
$V(\phi)$ is displaced at $\phi$=0. Then, if
$V(\phi)=\lambda \phi^{\nu}$ during inflation, we
have
\begin{equation}
N(\phi_{i} \rightarrow \phi_{f})=
-\int^{\phi_{f}}_{\phi_{i}}
\frac {3H^2(\phi)d\phi}{V^{\prime}(\phi)}=
-8\pi G\int^{\phi_{f}}_{\phi_{i}} \frac
{V(\phi)d\phi}{V^{\prime}(\phi)}=
\frac {4 \pi G}{\nu}(\phi^{2}_{i}-\phi^{2}_{f})~.
\label{eq:expefold}
\end{equation}
Assuming that $\phi_{i}\gg\phi_{f}$, this reduces
to $N(\phi)\approx(4 \pi G/\nu)\phi^2$.

\section{Coherent Oscillations of the Inflaton}
\label{sec:osci}

After the end of inflation at $t_f$, the term
$\ddot{\phi}$ takes over in (\ref{eq:reduce})
and $\phi$ starts performing coherent damped
oscillations about the global minimum of the
potential. The rate of energy density loss, due
to `friction', is given by
\begin{equation}
\dot{\rho}=\frac{d}{dt}\left(\frac{1}{2}
\dot{\phi}^2+V(\phi)\right)=-3H\dot{\phi}^2=
-3H(\rho+p)~,
\label{eq:damp}
\end{equation}
where $\rho=\dot{\phi}^2/2+V(\phi)$ and
$p=\dot{\phi}^2/2-V(\phi)$. Averaging $p$ over
one oscillation of $\phi$ and writing $\rho+p=
\gamma\rho$, we get $\rho\propto a^{-3\gamma}$
and $a(t)\propto t^{2/3\gamma}$ (see
Sect.\ref{subsec:friedmann}).

\par
The number $\gamma$ can be written as (assuming a
symmetric potential)
\begin{equation}
\gamma=\frac{\int^{T}_{0}\dot{\phi}^{2}dt}
{\int^{T}_{0}\rho dt}=
\frac{\int^{\phi_{{\rm{max}}}}_{0}
\dot{\phi}d\phi}{\int^{\phi_{{\rm{max}}}}_{0}
(\rho/\dot{\phi})d\phi}~,
\label{eq:gamma}
\end{equation}
where $T$ and $\phi_{{\rm{max}}}$ are the period
and the amplitude of the oscillation. From
$\rho=\dot{\phi}^2/2+V(\phi)=V_{{\rm{max}}}$, where
$V_{\rm max}$ is the maximal potential energy
density, we obtain $\dot{\phi}=
\sqrt{2(V_{{\rm{max}}}-V(\phi))}$. Substituting
this in (\ref{eq:gamma}) we get \cite{oscillation}
\begin{equation}
\gamma=\frac{2\int^{\phi_{{\rm{max}}}}_{0}
(1-V/V_{{\rm{max}}})^\frac{1}{2} d\phi}
{\int^{\phi_{{\rm{max}}}}_{0}
(1-V/V_{{\rm{max}}})^{-\frac{1}{2}}d\phi}~\cdot
\label{eq:gammafinal}
\end{equation}
For $V(\phi)=\lambda\phi^{\nu}$, we find $\gamma=
2\nu/(\nu+2)$ and, thus, $\rho\propto
a^{-6\nu/(\nu+2)}$ and $a(t)\propto
t^{(\nu+2)/3\nu}$. For $\nu=2$, in particular,
$\gamma=1$, $\rho\propto a^{-3}$, $a(t)\propto
t^{2/3}$ and $\phi$ behaves like pressureless
matter. This is not unexpected since a coherent
oscillating massive free field corresponds to a
distribution of static massive particles. For
$\nu$=4, we obtain $\gamma=4/3$,
$\rho\propto a^{-4}$, $a(t)\propto t^{1/2}$ and
the system resembles radiation. For $\nu = 6$, one
has $\gamma=3/2$, $\rho\propto a^{-9/2}$, $a(t)
\propto t^{4/9}$ and the expansion is slower (the
pressure is higher) than in radiation.

\section{Decay of the Inflaton}
\label{sec:decay}

Reintroducing the `decay term' $\Gamma_{\phi}
\dot{\phi}$, (\ref{eq:evolution}) can be written as
\begin{equation}
\dot{\rho}=\frac{d}{dt}\left(\frac{1}{2}\dot{\phi}^2
+V(\phi)\right)=-(3H+\Gamma_\phi)\dot{\phi}^2,
\label{eq:decay}
\end{equation}
which is solved \cite{reheat,oscillation} by
\begin{equation}
\rho(t)=\rho_{f} \left(\frac{a(t)}{a(t_{f})}
\right)^{-3 \gamma}{\rm{exp}}[-\gamma
\Gamma_{\phi}(t-t_f)]~,
\label{eq:rho}
\end{equation}
where $\rho_f$ is the energy density at $t_f$. The
second and third factors in the rhs of this equation
represent the dilution of the field energy due to the
expansion of the universe and the decay of $\phi$ to
`light' particles respectively.

\par
All pre-existing radiation (known as `old radiation')
was diluted by inflation, so the only radiation
present is the one produced by the decay of $\phi$
and is known as `new radiation'. Its energy density
satisfies \cite{reheat,oscillation} the equation
\begin{equation}
\dot{\rho}_{r}=-4H \rho_{r}+\gamma\Gamma_{\phi}\rho~,
\label{eq:newrad}
\end{equation}
where the first term in the rhs represents the
dilution of radiation due to the cosmological
expansion while the second one is the energy density
transfer from $\phi$ to radiation. Taking
$\rho_{r}(t_f)$=0, this equation gives
\cite{reheat,oscillation}
\begin{equation}
\rho_{r}(t)=\rho_{f}\left(\frac {a(t)}
{a(t_{f})}\right)^{-4}\int^{t}_{t_{f}}
\left(\frac{a(t^{\prime})} {a(t_{f})}\right)^{4-3
\gamma} e^{ -\gamma \Gamma_{\phi} (t^{\prime}-t_f)}
~\gamma\Gamma_{\phi} dt^{\prime}~.
\label{eq:rad}
\end{equation}
For $t_{f} \ll t_{d}$ and $\nu =2$, this expression
is approximated by
\begin{equation}
\rho_{r}(t)=\rho_{f}\left(\frac {t}{t_f}
\right)^{-\frac{8}{3}}\int^{t}_{0}\left(
\frac{t^{\prime}}{t_{f}}\right)^\frac{2}{3}
e^{-\Gamma_{\phi}t^{\prime}}dt^{\prime}~,
\label{eq:appr}
\end{equation}
which can be expanded as
\begin{equation}
\rho_{r}=\frac {3}{5}~\rho~\Gamma_{\phi}t\left[1+
\frac{3}{8}~\Gamma_{\phi}t+\frac
{9}{88}~(\Gamma_{\phi}t)^2+\cdots\right],
\label{eq:expand}
\end{equation}
with $\rho=\rho_{f} (t/t_{f})^{-2}{\rm{exp}}
(-\Gamma_{\phi}t)$ being the energy density of the
field $\phi$.

\par
The energy density of the `new radiation' grows
relative to the energy density of the oscillating
field and becomes essentially equal to it at a
cosmic time $t_{d}=\Gamma_{\phi}^{-1}$ as one
can deduce from (\ref{eq:expand}). After this time,
the universe enters into the radiation dominated
era and the normal big bang cosmology is recovered.
The temperature at $t_{d}$, $T_{r}(t_{d})$, is
historically called the `reheat' temperature
although no supercooling and subsequent reheating
of the universe actually takes place. Using
(\ref{eq:temptime}), we find that
\begin{equation}
T_{r}=\left(\frac {45}{16 \pi^{3}g_*}
\right)^\frac{1}{4}(\Gamma_{\phi}M_{P})^\frac{1}{2},
\label{eq:reheat}
\end{equation}
where $g_*$ is the effective number of degrees of
freedom. For $V(\phi)=\lambda\phi^{\nu}$, the total
expansion of the universe during the damped field
oscillations is
\begin{equation}
\frac{a(t_{d})}{a(t_{f})}=\left(
\frac{t_{d}}{t_{f}}\right)^{\frac{\nu+2}{3\nu}}.
\label{eq:expansion}
\end{equation}

\section{Density Perturbations from Inflation}
\label{sec:density}

Inflation not only homogenizes the universe but
also generates the density perturbations needed
for structure formation. To see this, we
introduce the notion of event horizon at $t$.
This includes all points with which we will
eventually communicate sending signals at $t$.
Its `instantaneous' radius is
\begin{equation}
d_{e}(t)=a(t)\int^{\infty}_{t}
\frac{dt^{\prime}}{a(t^{\prime})}~\cdot
\label{eq:event}
\end{equation}
This yields an infinite event horizon for
`matter' or `radiation'. For inflation,
however, we obtain $d_{e}(t)=H^{-1} <\infty$,
which varies slowly with $t$. Points, in our
event horizon at $t$, with which we can
communicate sending signals at $t$, are
eventually pulled away by the exponential
expansion and we cease to be able to
communicate with them emitting signals at
later times. We say that these points (and the
corresponding scales) crossed outside the event
horizon. Actually, the exponentially expanding
(de Sitter) space is like a black hole turned
inside out. Then, exactly as in a black hole,
there are quantum fluctuations of the `thermal
type' governed by the Hawking temperature
\cite{hawking,gibbons} $T_{H}=H/2\pi$. It turns out
\cite{bunch,vilenkin} that the quantum fluctuations
of all massless fields (the inflaton is nearly
massless due to the flatness of the potential) are
$\delta\phi=T_{H}$. These fluctuations of $\phi$
lead to energy density perturbations $\delta\rho=
V^{\prime}(\phi)\delta\phi$. As the scale of this
perturbations crosses outside the event horizon,
they become \cite{fischler} classical metric
perturbations.

\par
It has been shown \cite{zeta} that the quantity
$\zeta\approx\delta\rho/(\rho+p)$ remains constant
outside the event horizon. Thus, the density
perturbation at any present physical (`comoving')
scale $\ell$, $(\delta\rho/\rho)_{\ell}$, when this
scale crosses inside the post-inflationary particle
horizon ($p$=0 at this instance) can be related to
the value of $\zeta$ when the same scale crossed
outside the inflationary event horizon (at
$\ell\sim H^{-1}$). This latter value of $\zeta$ is
found, using (\ref{eq:infeq}), to be
\begin{equation}
\zeta\mid_{\ell\sim H^{-1}}=\left(\frac
{\delta\rho}{\dot{\phi}^2}\right)_{\ell\sim
H^{-1}}=\left(\frac{V^{\prime}(\phi)H(\phi)}{2\pi
\dot{\phi}^2}\right)_{\ell\sim H^{-1}}=-\left(
\frac {9 H^{3}(\phi)} {2\pi V^{\prime}(\phi)}
\right)_{\ell\sim H^{-1}}\cdot
\label{eq:zeta}
\end{equation}
Taking into account an extra 2/5 factor from the
fact that the universe is matter dominated when
the scale $\ell$ re-enters the horizon, we obtain
\begin{equation}
\left(\frac{\delta\rho}{\rho}\right)_{\ell}=
\frac{16\sqrt{6\pi}}{5}~\frac {V^\frac{3}{2}
(\phi_{\ell})}{M^{3}_{P} V^{\prime}(\phi_{\ell})}
~\cdot
\label{eq:deltarho}
\end{equation}

\par
The calculation of $\phi_{\ell}$, the value of $\phi$
when the `comoving' scale $\ell$ crossed outside the
event horizon, goes as follows. A `comoving' (present
physical) scale $\ell$, at $T_r$, was equal to
$\ell(a(t_{d})/a(t_{0}))=\ell(T_{0}/T_{r})$. Its
magnitude at $t_{f}$ was equal to
$\ell(T_{0}/T_{r})(a(t_{f})/a(t_{d}))=
\ell(T_{0}/T_{r})(t_{f}/t_{d})^{(\nu+2)/3 \nu}
\equiv\ell_{{\rm{phys}}}(t_{f})$, where the potential
$V(\phi)=\lambda\phi^{\nu}$ was assumed. The scale
$\ell$, when it crossed outside the inflationary
event horizon, was equal to $H^{-1}(\phi_{\ell})$. We,
thus, obtain
\begin{equation}
H^{-1}(\phi_{\ell}) e^{N(\phi_{\ell})} =
\ell_{{\rm{phys}}}(t_{f})~,
\label{eq:lphys}
\end{equation}
which gives $\phi_{\ell}$ and, thus, $N(\phi_{\ell})
\equiv N_{\ell}$, the number of e-foldings the scale
$\ell$ suffered during inflation. In particular, the
number of e-foldings suffered by our present horizon
$\ell=2H_{0}^{-1}\sim 10^4~{\rm Mpc}$ turns out to
be $N_{Q}\approx 50-60$.

\par
Taking $V(\phi)=\lambda \phi^4$, (\ref{eq:expefold}),
(\ref{eq:deltarho}) and (\ref{eq:lphys}) give
\begin{equation}
\left(\frac{\delta\rho}{\rho}\right)_{\ell}=
\frac{4\sqrt{6\pi}}{5}\lambda^\frac{1}{2}\left(
\frac{\phi_{\ell}}{M_{P}}\right)^3=
\frac{4 \sqrt{6 \pi}}{5}\lambda^\frac{1}{2}
\left(\frac{N_{\ell}}{\pi}\right)^\frac{3}{2}\cdot
\label{eq:nl}
\end{equation}
From the result of COBE \cite{cobe},
$(\delta\rho/\rho)_{Q}\approx 6\times 10^{-5}$,
one can then deduce that $\lambda\approx 6\times
10^{-14}$ for $N_{Q}\approx 55$. We thus see
that the inflaton must be a very weakly coupled
field. In non-SUSY GUTs, the inflaton is
necessarily gauge singlet since otherwise radiative
corrections will make it strongly coupled. This is
not so satisfactory since it forces us to introduce
an otherwise unmotivated very weakly coupled gauge
singlet. In SUSY GUTs, however, the inflaton could
be identified \cite{nonsinglet} with a conjugate
pair of gauge non-singlet fields $\bar{\phi}$,
$\phi$ already present in the theory and
causing the gauge symmetry breaking. Absence
of strong radiative corrections from gauge
interactions is guaranteed by the mutual
cancellation of the D-terms of these fields.

\par
The spectrum of density perturbations can be
analyzed. For $V(\phi)=\lambda\phi^{\nu}$, we
find $(\delta\rho/\rho)_{\ell}\propto
\phi_{\ell}^{(\nu+2)/2}$ which, together with
$N(\phi_{\ell})\propto\phi_{\ell}^{2}$ (see
(\ref{eq:expefold})), gives
\begin{equation}
\left(\frac{\delta\rho}{\rho}\right)_{\ell}=\left(
\frac{\delta\rho}{\rho}\right)_{Q}\left(
\frac{N_{\ell}}{N_{Q}}\right)^{\frac{\nu+2}{4}}.
\label{eq:spectrum}
\end{equation}
The scale $\ell$ divided by the size of our present
horizon ($2H_{0}^{-1}\sim 10^4~{\rm{Mpc}}$) should
equal ${\rm exp}(N_{\ell}-N_{Q})$. This gives
$N_{\ell}/N_{Q}=1+\ln(\ell/2H_{0}^{-1})^{1/N_{Q}}$
which expanded around $\ell=2H_{0}^{-1}$
and substituted in (\ref{eq:spectrum}) yields
\begin{equation}
\left(\frac{\delta \rho}{\rho}\right)_{\ell}\approx
\left(\frac{\delta \rho}{\rho}\right)_{Q}\left(
\frac{\ell}{2H_{0}^{-1}}\right)^{\alpha_{s}},
\label{eq:alphas}
\end{equation}
with $\alpha_{s}=(\nu+2)/4N_{Q}$. For $\nu=4$,
$\alpha_{s}\approx 0.03$ and, thus, the density
perturbations are essentially scale independent.
The customarily used spectral index
$n=1-2\alpha_{s}$ is about 0.94 in this case.

\section{Temperature Fluctuations}
\label{sec:temperature}

The density inhomogeneities produce temperature
fluctuations in the CMBR. For angles $\theta
\stackrel{_{>}}{_{\sim }}2^{o}$, the dominant effect
is the scalar Sachs-Wolfe \cite{sachswolfe} effect.
Density perturbations on the `last scattering
surface' cause scalar gravitational potential
fluctuations, which then produce temperature
fluctuations in the CMBR. The reason is that regions
with a deep gravitational potential will cause the
photons to lose energy as they climb up the well and,
thus, these regions will appear cooler.

\par
Analyzing the temperature fluctuations from the
scalar Sachs-Wolfe effect in spherical harmonics,
we obtain the corresponding quadrupole
anisotropy:
\begin{equation}
\left(\frac{\delta T}{T}\right)_{Q-S}=\left(
\frac{32 \pi}{45}\right)^\frac{1}{2}
\frac{V^\frac{3}{2}(\phi_{\ell})}{M^{3}_{P}
V^{\prime}(\phi_{\ell})}~\cdot
\label{eq:quadrupole}
\end{equation}
For $V(\phi)=\lambda\phi^{\nu}$, this becomes
\begin{equation}
\left(\frac{\delta T}{T}\right)_{Q-S}=\left(
\frac{32\pi}{45}\right)^\frac{1}{2}
\frac{\lambda^\frac{1}{2}
\phi_{\ell}^{\frac{\nu+2}{2}}}{\nu M^{3}_{P}}
=\left(\frac{32 \pi}{45}\right)^\frac{1}{2}
\frac{\lambda^\frac{1}{2}}{\nu M^{3}_{P}}
\left(\frac{\nu M^{2}_{P}}{4\pi}
\right)^{\frac{\nu+2}{4}}
N_{\ell}^{\frac{\nu+2}{4}}.
\label{eq:anisotropy}
\end{equation}
Comparing this with the COBE \cite{cobe} result,
$(\delta T/T)_{Q}\approx 6.6\times 10^{-6}$, we
obtain $\lambda\approx 6\times 10^{-14}$ for
$\nu=4$ and number of e-foldings suffered by our
present horizon scale during the inflationary
phase $N_{Q}\approx 55$. The `tensor' fluctuations
\cite{tensor}, which generally can also exist in
the temperature of the CMBR, turn out to be
negligible in all cases considered here.

\section{Hybrid Inflation}
\label{sec:hybrid}

\subsection{The non-Supersymmetric Version}
\label{subsec:nonsusy}

The main disadvantage of inflationary scenarios
such as the `new' \cite{new} or `chaotic'
\cite{chaotic} ones is that they require tiny
parameters in order to reproduce the results of
COBE \cite{cobe}. This has led Linde
\cite{hybrid} to propose, in the context of
non-SUSY GUTs, hybrid inflation which uses two
real scalar fields $\chi$ and $\sigma$ instead
of one. $\chi$ provides the `vacuum' energy
density driving inflation, while $\sigma$ is
the slowly varying field during inflation. This
allows us to reproduce the COBE results with
`natural' (not too small) values of the
parameters.

\par
The scalar potential utilized by Linde is
\begin{equation}
V(\chi,\sigma)=\kappa^2 \left(M^2-\frac{\chi^2}{4}
\right)^2+\frac{\lambda^2\chi^2 \sigma^2}{4}+
\frac{m^2\sigma^2}{2}~,
\label{eq:lindepot}
\end{equation}
where $\kappa,~\lambda>0$ are dimensionless
constants and $M$, $m$ mass parameters. The
vacua lie at $\langle\chi\rangle=\pm 2M$,
$\langle\sigma\rangle=0$. For $m$=0, $V$ has a flat
direction at $\chi=0$, where $V=\kappa^2M^4$ and
the ${\rm mass}^2$ of $\chi$ is $m^2_\chi=
-\kappa^2M^2+\lambda^2\sigma^2/2$. So, for $\chi=0$
and $\vert\sigma\vert>\sigma_c=\sqrt{2}\kappa M/
\lambda$, we obtain a flat valley of minima. For
$m\neq 0$, the valley acquires a slope and the
system can inflate rolling down this valley.

\par
The $\epsilon$ and $\eta$ criteria (see
(\ref{eq:src})) imply that inflation continues
until $\sigma$ reaches $\sigma_c$, where it
terminates abruptly. It is followed by a
`waterfall', i.e., a sudden entrance into an
oscillatory phase about a global minimum. Since
the system can fall into either of the two minima
with equal probability, topological defects
(monopoles, cosmic strings or walls) are copiously
produced \cite{smooth} if they are predicted by the
particular GUT employed. So, if the underlying GUT
gauge symmetry breaking (by $\langle\chi\rangle$)
leads to the existence of monopoles or walls, we
encounter a catastrophe.

\par
The onset of hybrid inflation requires \cite{onset}
that, at $t\sim H^{-1}$, $H$ being the inflationary
Hubble parameter, a region exists with size
$\stackrel{_{>}}{_{\sim}}H^{-1}$, where $\chi$ and
$\sigma$ are almost uniform with negligible kinetic
energies and values close to the bottom of the
valley of minima. Such a region, at $t_P$, would
have been much larger than the Planck length
$\ell_P$ and it is, thus, difficult to imagine how
it could be so homogeneous. Moreover, as it has
been argued \cite{initial}, the initial values (at
$t_P$) of the fields in this region must be
strongly restricted in order to obtain adequate
inflation. Several possible solutions to this
problem of initial conditions for hybrid inflation
have been proposed (see e.g.,
\cite{double,sugra,costas}).

\par
The quadrupole anisotropy of the CMBR produced
during hybrid inflation can be estimated, using
(\ref{eq:quadrupole}), to be
\begin{equation}
\left(\frac{\delta T}{T}\right)_{Q}\approx
\left(\frac{16\pi}{45}\right)^{\frac{1}{2}}
\frac{\lambda\kappa^2M^5}{M^3_Pm^2}~\cdot
\label{eq:lindetemp}
\end{equation}
The COBE \cite{cobe} result, $(\delta T/T)_{Q}
\approx 6.6\times 10^{-6}$, can then be
reproduced with $M\approx 2.86\times 10^{16}
~{\rm GeV}$, the SUSY GUT vev, and $m\approx 1.3
~\kappa\sqrt{\lambda}\times 10^{15}~{\rm GeV}$.
Note that $m\sim 10^{12}~{\rm GeV}$ for $\kappa$,
$\lambda\sim 10^{-2}$.

\subsection{The Supersymmetric Version}
\label{subsec:susy}

Hybrid inflation is \cite{lyth} `tailor made' for
globally SUSY GUTs except that an intermediate
scale mass for $\sigma$ cannot be obtained.
Actually, all scalars acquire masses $\sim m_{3/2}
\sim 1~{\rm TeV}$ (the gravitino mass) from soft
SUSY breaking.

\par
Let us consider the renormalizable superpotential
\begin{equation}
W=\kappa S(-M^2+\bar{\phi}\phi)~,
\label{eq:superpot}
\end {equation}
where $\bar{\phi}$, $\phi$ is a  pair of $G_{S}$
singlet left handed superfields belonging to
conjugate representations of $G$ and reducing its
rank by their vevs, and $S$ is a gauge singlet
left handed superfield. $\kappa$ and $M$ ($\sim
10^{16}~{\rm GeV}$) are made positive by field
redefinitions. The vanishing of the F-term $F_S$
gives
$\langle\bar{\phi}\rangle\langle\phi\rangle=M^2$,
and the D-terms vanish for $\vert\langle\bar{\phi}
\rangle\vert=\vert\langle\phi\rangle\vert$. So,
the SUSY vacua lie at $\langle\bar{\phi}
\rangle^*=\langle\phi\rangle=\pm M$ and $\langle
S\rangle=0$ (from $F_{\bar{\phi}}=F_{\phi}=0$).
Thus, $W$ leads to the breaking of $G$.

\par
$W$ also gives rise to hybrid inflation.
The potential derived from $W$ is
\begin{equation}
V(\bar{\phi},\phi,S)=\kappa^2\vert M^2-
\bar{\phi}\phi\vert^2+
\kappa^2\vert S\vert^2(\vert\bar{\phi}\vert^2+
\vert\phi\vert^2)+{\rm{D-terms}}~.
\label{eq:hybpot}
\end{equation}
D-flatness implies $\bar{\phi}^*=e^{i\theta}\phi$.
We take $\theta=0$, so that the SUSY vacua are
contained. $W$ has a $U(1)_R$ R-symmetry:
$\bar{\phi}\phi\to\bar{\phi}\phi$,
$S\to e^{i\alpha}S$, $W\to e^{i\alpha}W$. Actually,
$W$ is the most general renormalizable
superpotential allowed by $G$ and $U(1)_R$.
Bringing $\bar{\phi}$, $\phi$, $S$ on the real
axis by $G$ and $U(1)_R$ transformations, we write
$\bar{\phi}=\phi\equiv\chi/2$,
$S\equiv\sigma/\sqrt{2}$ where $\chi$, $\sigma$
are normalized real scalar fields. $V$ then takes
the form in (\ref{eq:lindepot}) with
$\kappa=\lambda$ and $m=0$. So, Linde's potential
is almost obtainable from SUSY GUTs but without
the mass term of $\sigma$.

\par
SUSY breaking by the `vacuum' energy density
$\kappa^2M^4$ on the inflationary valley
($\bar{\phi}=\phi=0$, $\vert S\vert>S_{c}\equiv M$)
causes a mass splitting in the supermultiplets
$\bar{\phi}$, $\phi$. We obtain a Dirac fermion
with ${\rm mass}^2=\kappa^2\vert S\vert^2$ and two
complex scalars with ${\rm mass}^2=\kappa^2\vert S
\vert^2\pm\kappa^2M^2$. This leads \cite{dss} to
one-loop corrections to $V$ on the valley via the
Coleman-Weinberg formula \cite{cw}:
\begin{equation}
\Delta V=\frac{1}{64\pi^2}\sum_i(-)^{F_i}\ M_i^4
\ln\frac{M_i^2}{\Lambda^2}~,
\label{eq:deltav}
\end{equation}
where the sum extends over all helicity states $i$,
with fermion number $F_i$ and ${\rm mass}^2=M_i^2$,
and $\Lambda$ is a renormalization scale. We find
that $\Delta V(\vert S\vert)$ is
\begin{equation}
\kappa^2 M^4~{\kappa^2N\over 32\pi^2}\left(
2\ln{\kappa^2\vert S\vert^2\over\Lambda^2}
+(z+1)^{2}\ln(1+z^{-1})+(z-1)^{2}\ln(1-z^{-1})
\right),
\label{eq:rc}
\end{equation}
where $z=x^2=\vert S\vert^2/M^2$ and $N$ is the
dimensionality of the representations to which
$\bar{\phi}$, $\phi$ belong. These radiative
corrections generate the necessary slope on the
inflationary valley. Note that the slope is
$\Lambda$-independent.

\par
From (\ref{eq:expefold}), (\ref{eq:quadrupole})
and (\ref{eq:rc}), we find the quadrupole
anisotropy of the CMBR:
\begin{equation}
\left(\frac{\delta T}{T}\right)_{Q}\approx
\frac{8\pi}{\sqrt{N}}\left(\frac{N_{Q}}{45}
\right)^{\frac{1}{2}}\left(\frac{M}{M_{P}}
\right)^2x_Q^{-1}y_Q^{-1}\Lambda(x_Q^2)^{-1},
\label{eq:qa}
\end{equation}
with
\begin{equation}
\Lambda(z)=(z+1)\ln(1+z^{-1})+(z-1)\ln(1-z^{-1})~,
\label{eq:lambda}
\end{equation}
\begin{equation}
y_Q^2=\int_1^{x_Q^2}\frac{dz}{z}\Lambda(z)^{-1},
~y_Q\geq 0~.
\label{eq:yq}
\end{equation}
Here, $x_Q=\vert S_Q\vert/M$, with $S_Q$ being
the value of $S$ when our present horizon crossed
outside the inflationary horizon. Finally, from
(\ref{eq:rc}), one finds
\begin{equation}
\kappa\approx\frac{8\pi^{\frac{3}{2}}}{\sqrt{NN_Q}}
~y_Q~\frac{M}{M_{P}}~\cdot \label{eq:kappa}
\end{equation}

\par
The slow roll conditions for SUSY hybrid inflation
are $\epsilon,\vert\eta\vert\leq 1$, where
\begin{equation}
\epsilon=\left(\frac{\kappa^2M_P}{16\pi^2M}\right)^2
\frac{N^2x^2}{8\pi}\Lambda(x^2)^2,
\label{eq:epsilon}
\end{equation}
\begin{equation}
\eta=\left(\frac{\kappa M_P}{4\pi M}\right)^2
\frac{N}{8\pi}\left((3z+1)\ln(1+z^{-1})+
(3z-1)\ln(1-z^{-1})\right).
\label{eq:eta}
\end{equation}
These conditions are violated only `infinitesimally'
close to the critical point ($x=1$). So, inflation
continues until this point, where the `waterfall'
occurs.

\par
Using COBE \cite{cobe} and eliminating $x_Q$ between
(\ref{eq:qa}) and (\ref{eq:kappa}), we obtain $M$ as
a function of $\kappa$. The maximal $M$ which can be
achieved is $\approx 10^{16}~{\rm GeV}$ (for $N=8$,
$N_Q\approx 55)$ and, although somewhat smaller than
the SUSY GUT vev, is quite close to it. As an
example, take $\kappa=4\times 10^{-3}$ which gives
$M\approx 9.57\times 10^{15}~{\rm GeV}$,
$x_Q\approx2.633$, $y_Q\approx 2.42$. The slow roll
conditions are violated at $x-1\approx 7.23\times
10^{-5}$, where $\eta=-1$ ($\epsilon\approx 8.17
\times 10^{-8}$ at $x=1$). The spectral index
$n=1-6\epsilon+2\eta$ \cite{liddle} is about 0.985.

\par
SUSY hybrid inflation is considered `natural' for
the following reasons:
\begin{list}
\setlength{\rightmargin=0cm}{\leftmargin=0cm}
\item[{\bf i.}] There is no need of tiny coupling
constants ($\kappa\sim 10^{-3}$).
\vspace{.25cm}
\item[{\bf ii.}] $W$ in (\ref{eq:superpot}) has
the most general renormalizable form allowed by
$G$ and $U(1)_R$. The coexistence of the $S$ and
$S\bar{\phi}\phi$ terms implies that
$\bar{\phi}\phi$ is `neutral' under all symmetries
and, thus, all the non-renormalizable terms of
the form $S(\bar{\phi}\phi)^n$, $n\geq 2$, are
also allowed \cite{jean}. The leading term of this
type $S(\bar{\phi}\phi)^2$, if its dimensionless
coefficient is of order unity, can be comparable
to $S\bar{\phi}\phi$ (recall that $\kappa\sim
10^{-3}$) and, thus, play a role in inflation (see
Sect.\ref{sec:extensions}). $U(1)_R$ guarantees
the linearity of $W$ in $S$ to all orders
excluding terms such as $S^2$ which could generate
an inflaton mass $\stackrel{_{>}}{_{\sim }}H$ and
ruin inflation by violating the slow roll
conditions.
\vspace{.25cm}
\item[{\bf iii.}] SUSY guarantees that the
radiative corrections do not ruin \cite{nonsinglet}
inflation, but rather provide \cite{dss} the
necessary slope on the inflationary path.
\vspace{.25cm}
\item[{\bf iv.}] Supergravity corrections can be
brought under control leaving inflation intact
\cite{sugra,costas,lss}.
\end{list}
In summary, for all these reasons, we consider
SUSY hybrid inflation (with its extensions) as an
extremely `natural' inflationary scenario.

\section{Extensions of Supersymmetric Hybrid
Inflation}
\label{sec:extensions}

Applying (SUSY) hybrid inflation to higher GUT
gauge groups predicting monopoles, we encounter
the following problem. Inflation is terminated
abruptly as the system reaches the critical point
and is followed by the `waterfall' regime during
which the scalar fields $\bar\phi$, $\phi$
develop their vevs starting from zero and the
spontaneous breaking of the GUT gauge symmetry
occurs. The fields $\bar\phi$, $\phi$ can
end up at any point of the vacuum manifold with
equal probability and, thus, monopoles are
copiously produced \cite{smooth} via the Kibble
mechanism \cite{kibble} leading to a disaster.

\par
One of the simplest GUTs predicting monopoles is
the Pati-Salam (PS) model \cite{ps} with gauge
group $G_{PS}=SU(4)_c\times SU(2)_L \times
SU(2)_R$. These monopoles carry two units of
`Dirac' magnetic charge \cite{magg}. We will
present solutions \cite{smooth,jean} of the
monopole problem of hybrid inflation within the
SUSY PS model, although our mechanisms can be
extended to other gauge groups such as the
`trinification' group $SU(3)_c\times SU(3)_L
\times SU(3)_R$, which predicts
\cite{trinification} monopoles with
triple `Dirac' charge.

\subsection{Shifted Hybrid Inflation}
\label{subsec:shifted}

One idea \cite{jean} for solving the monopole
problem is to include into the standard
superpotential for hybrid inflation (in
(\ref{eq:superpot})) the leading
non-renormalizable term, which, as explained,
cannot be excluded. If its dimensionless
coefficient is of order unity, this term
competes with the trilinear term of the
standard superpotential (with coefficient
$\sim 10^{-3}$). A totally new picture then
emerges. There appears a non-trivial flat
direction along which $G_{PS}$ is broken
with the appropriate Higgs fields acquiring
constant values. This `shifted' flat
direction acquires a slope again from
radiative corrections \cite{dss} and can be
used for inflation. The end of inflation is
again abrupt followed by a `waterfall' but
no monopoles are formed since $G_{PS}$ is
already broken during inflation.

\par
The spontaneous breaking of the gauge group
$G_{PS}$ to $G_S$ is achieved via the vevs of
a conjugate pair of Higgs superfields
\begin{eqnarray}
\bar{H}^c &=& (4,1,2)\equiv
\left(\begin{array}{cccc}
\bar{u}^c_H & \bar{u}^c_H &
\bar{u}^c_H & \bar{\nu}_H^c\\
\bar{d}^c_H & \bar{d}^c_H &
\bar{d}^c_H & \bar{e}^c_H
\end{array}\right),
\nonumber\\
H^c &=& (\bar{4},1,2)\equiv
\left(\begin{array}{cccc}
u^c_H & u^c_H & u^c_H & \nu_H^c\\
d^c_H & d^c_H & d^c_H & e^c_H
\end{array}\right),
\label{eq:higgs}
\end{eqnarray}
in the $\bar{\nu}_H^c$, $\nu_H^c$ directions. The
relevant part of the superpotential, which includes
the leading non-renormalizable term, is
\begin{equation}
\delta W=\kappa S(-M^2+\bar{H}^c H^c)-
\beta\frac{S(\bar{H}^c
H^c)^2}{M_S^2}~,
\label{eq:susyinfl}
\end{equation}
where $M_S\approx 5\times 10^{17}~{\rm GeV}$ is the
string scale and $\beta$ is taken positive for
simplicity. D-flatness implies that
$\bar{H}^{c} \,^{*}=e^{i\theta}H^c$. We restrict
ourselves to the direction with $\theta=0$
($\bar{H}^{c} \,^{*}=H^c$) containing the `shifted'
inflationary path (see below). The scalar potential
derived from $\delta W$ then takes the form
\begin{equation}
V=\left[\kappa(\vert H^c\vert^2-M^2)-
\beta\frac{\vert H^c\vert^4}{M_S^2}\right]^2+
2\kappa^2\vert S\vert^2 \vert H^c\vert^2\left[1-
\frac{2\beta}{\kappa M_S^2}\vert H^c\vert^2\right]^2.
\label{eq:inflpot}
\end{equation}
Defining the dimensionless variables $w=\vert S
\vert/M$, $y=\vert H^c\vert/M$, we obtain
\begin{equation}
\tilde{V}=\frac{V}{\kappa^2M^4}=(y^2-1-\xi y^4)^2+
2w^2y^2(1-2\xi y^2)^2,
\label{eq:vtilde}
\end{equation}
where $\xi=\beta M^2/\kappa M_S^2$. This potential
is a simple extension of the standard potential for
SUSY hybrid inflation (which corresponds to
$\xi=0$).

\par
For constant $w$ (or $|S|$), $\tilde V$ in
(\ref{eq:vtilde}) has extrema at
\begin{equation}
y_1=0,~y_2=\frac{1}{\sqrt{2\xi}},~y_{3\pm}=
\frac{1}{\sqrt{2\xi}}\sqrt{(1-6\xi w^2)\pm
\sqrt{(1-6\xi w^2)^2-4\xi(1-w^2)}}.
\label{eq:extrema}
\end{equation}
The first two extrema (at $y_1$, $y_2$) are
$|S|$-independent and, thus, correspond to
flat directions, the trivial one at $y_1=0$ with
$\tilde{V}_1=1$, and the `shifted' one at
$y_2=1/\sqrt{2\xi}={\rm constant}$ with
$\tilde{V}_2=(1/4\xi-1)^2$, which we will use as
inflationary path. The trivial trajectory is a
valley of minima for $w>1$, while the `shifted'
one for $w>w_0=(1/8\xi-1/2)^{1/2}$, which is its
critical point. We take $\xi<1/4$, so that
$w_0>0$ and the `shifted' path is destabilized
before $w$ reaches zero. The extrema at
$y_{3\pm}$, which are $|S|$-dependent and
non-flat, do not exist for all values of $w$ and
$\xi$, since the expressions under the square
roots in (\ref{eq:extrema}) are not always
non-negative. These two extrema, at $w=0$, become
SUSY vacua. The relevant SUSY vacuum (see below)
corresponds to $y_{3-}(w=0)$ and, thus, the
common vev $v_0$ of $\bar{H}^{c}$, $H^c$ is
\begin{equation}
\left(\frac{v_0}{M}\right)^2=
\frac{1}{2\xi}(1-\sqrt{1-4\xi})~.
\label{eq:v0}
\end{equation}

\par
We will now discuss the structure of $\tilde{V}$
and the inflationary history for $1/6<\xi<1/4$.
For fixed $w>1$, there exist two local minima at
$y_1=0$ and $y_2=1/\sqrt{2\xi}$, which has lower
potential energy density, and a local maximum at
$y_{3+}$ between the minima. As $w$ becomes
smaller than unity, the extremum at $y_1$ turns
into a local maximum, while the extremum at
$y_{3+}$ disappears. The system then falls into
the `shifted' path in case it had started at
$y_1=0$. As we further decrease $w$ below
$(2-\sqrt{36\xi-5})^{1/2}/3\sqrt{2\xi}$, a pair
of new extrema, a local minimum at $y_{3-}$ and
a local maximum at $y_{3+}$, are created between
$y_1$ and $y_2$. As $w$ crosses
$(1/8\xi-1/2)^{1/2}$, the local maximum at
$y_{3+}$ crosses $y_2$ becoming a local minimum.
At the same time, the local minimum at $y_2$
turns into a local maximum and inflation
ends with the system falling into the local
minimum at $y_{3-}$ which, at $w=0$, becomes the
SUSY vacuum.

\par
We see that, no matter where the system starts
from, it passes from the `shifted' path, where
the relevant part of inflation takes place. So,
$G_{PS}$ is broken during inflation and no
monopoles are produced at the `waterfall'.

\par
After inflation, the system could fall into
the minimum at $y_{3+}$ instead of the one
at $y_{3-}$. This, however, does not happen
since in the last e-folding or so the barrier
between the minima at $y_{3-}$ and $y_2$ is
considerably reduced and the decay of the
`false vacuum' at $y_2$ to the minimum at
$y_{3-}$ is completed within a fraction of
an e-folding before the $y_{3+}$ minimum
even appears.

\par
The only mass splitting within supermultiplets
on the `shifted' path appears \cite{jean}
between one Majorana fermion in the direction
$(\bar{\nu}_H^c+\nu_H^c)/\sqrt{2}$ with $m^2=
4\kappa^2\vert S\vert^2$ and two real scalars
$\mathrm{Re}(\delta\bar{\nu}^c_H+\delta
\nu^c_H)$ and $\mathrm{Im}(\delta
\bar{\nu}^c_H+\delta\nu^c_H)$ with
$m_{\pm}^2=4\kappa^2\vert S\vert^2\mp 2
\kappa^2m^2$. Here, $m=M(1/4\xi-1)^{1/2}$
and $\delta\bar{\nu}^c_H=\bar{\nu}^c_H-v$,
$\delta\nu^c_H=\nu^c_H-v$ where
$v=(\kappa M_S^2/2\beta)^{1/2}$ is the
value of $\bar{H}^c$, $H^c$ on the path.

\par
The radiative corrections on the `shifted'
path can be constructed and
$(\delta T/T)_Q$ and $\kappa$ can be
evaluated. We find the same formulas as in
(\ref{eq:qa}) and (\ref{eq:kappa}) with
$N=2$ and $N=4$ respectively and $M$
generally replaced by $m$. COBE \cite{cobe}
can be reproduced, for instance, with
$\kappa\approx 4\times 10^{-3}$,
corresponding to $\xi=1/5$, $v_0\approx 1.7
\times 10^{16}~{\rm GeV}$ ($N_Q\approx 55$,
$\beta=1$). The scales $M\approx 1.45\times
10^{16}~{\rm GeV}$, $m\approx 7.23\times
10^{15}~{\rm GeV}$, the inflaton mass
$m_{\mathrm{infl}}\approx 4.1\times 10^{13}
~{\rm GeV}$ and the `inflationary scale',
which characterizes the inflationary `vacuum'
energy density, $v_{\rm infl}=\kappa^{1/2}m
\approx 4.57\times 10^{14}~{\rm GeV}$. The
spectral index $n=0.954$.

\subsection{Smooth Hybrid Inflation}
\label{subsec:smooth}

\par
An alternative solution to the monopole problem
of hybrid inflation has been proposed in
\cite{smooth}. We will present it here within
the SUSY PS model of Sect.\ref{subsec:shifted},
although it can be
applied to other semi-simple gauge groups too.
The idea is to impose an extra $Z_2$ symmetry
under which $H^c\rightarrow -H^c$. The whole
structure of the model remains unchanged except
that now only even powers of the combination
$\bar{H}^cH^c$ are allowed in the superpotential
terms.

\par
The inflationary superpotential in
(\ref{eq:susyinfl}) becomes
\begin{equation}
\delta W=S\left(-\mu^2+\frac{(\bar{H}^cH^c)^2}
{M_S^2}\right),
\label{eq:smoothsuper}
\end{equation}
where we absorbed the dimensionless parameters
$\kappa$, $\beta$ in $\mu$, $M_S$. The resulting
scalar potential $V$ is then given by
\begin{equation}
\tilde{V}=\frac{V}{\mu^4}=(1-\tilde\chi^4)^2+
16\tilde\sigma^2\tilde\chi^6,
\label{eq:smoothpot}
\end{equation}
where we used the dimensionless fields
$\tilde\chi=\chi/2(\mu M_S)^{1/2}$, $\tilde\sigma
=\sigma/2(\mu M_S)^{1/2}$ with $\chi$, $\sigma$
being normalized real scalar fields defined by
$\bar{\nu}_H^c=\nu_H^c=\chi/2$,
$S=\sigma/\sqrt{2}$ after rotating $\bar{\nu}_H^c$,
$\nu_H^c$, $S$ to the real axis.

\par
The emerging picture is completely different. The
flat direction at $\tilde\chi=0$ is now a local
maximum with respect to $\tilde\chi$ for all values
of $\tilde\sigma$, and two new symmetric valleys of
minima appear \cite{smooth,shi} at
\begin{equation}
\tilde\chi=\pm\sqrt{6}\tilde\sigma\left[\left(1+
\frac{1}{36\tilde\sigma^4}\right)^{\frac{1}{2}}-1
\right]^{\frac{1}{2}}.
\label{eq:smoothvalley}
\end{equation}
They contain the SUSY vacua lying at $\tilde\chi=
\pm 1$, $\tilde\sigma=0$ and possess a slope
already at the classical level. So, in this case,
there is no need of radiative corrections for
driving the inflaton. The potential on these
paths is
\cite{smooth,shi}
\begin{equation}
\tilde{V}=48\tilde\sigma^4
\left[72\tilde\sigma^4\left(1+
\frac{1}{36\tilde\sigma^4}\right)
\left(\left(1+\frac{1}{36\tilde\sigma^4}
\right)^{\frac{1}{2}}-1\right)-1\right].
\label{eq:smoothV}
\end{equation}
The system follows a particular inflationary path
and ends up at a particular point of the vacuum
manifold leading to no production of monopoles.

\par
The end of inflation is not abrupt since the
inflationary path is stable with respect to
$\tilde\chi$ for all $\tilde\sigma$'s. It is
determined by using the $\epsilon$ and $\eta$
criteria.

\par
This model allows us to take the vev $v_0=
(\mu M_S)^{1/2}$ of $\bar{H}^c$, $H^c$
equal to the SUSY GUT vev. COBE \cite{cobe}
then yields $M_S\approx 4.39\times 10^{17}
~{\rm GeV}$ and $\mu\approx 1.86\times 10^{15}
~{\rm GeV}$ for $N_Q\approx 57$. Inflation ends
at $\sigma=\sigma_0\approx 1.34\times 10^{17}
~{\rm GeV}$, while our present horizon crosses
outside the inflationary horizon at $\sigma=
\sigma_Q\approx 2.71\times 10^{17}~{\rm GeV}$.
Finally, $m_{\rm infl}=2\sqrt{2}\mu^2/v_0
\approx 3.42\times 10^{14}~{\rm GeV}$.

\section{Conclusions}
\label{sec:conclusions}

\par
We summarized the shortcomings of SBB and
their resolution by inflation, which suggests
that the universe underwent a period of
exponential expansion. This may have happened
during the GUT phase transition at which the
relevant Higgs field was displaced from the
vacuum. This field (inflaton) could then, for
some time, roll slowly towards the vacuum
providing an almost constant `vacuum' energy
density. Inflation generates the density
perturbations needed for the large scale
structure of the universe and the temperature
fluctuations of the CMBR. After
the end of inflation, the inflaton performs
damped oscillations about the vacuum, decays
and `reheats' the universe.

\par
The early inflationary models required tiny
parameters. This problem was solved by hybrid
inflation which uses two real scalar fields. One
of them provides the `vacuum' energy density for
inflation while the other one is the slowly
rolling field. Hybrid inflation arises `naturally'
in many SUSY GUTs, but leads to a disastrous
overproduction of monopoles. We constructed two
extensions of SUSY hybrid inflation which do not
suffer from this problem.

\section*{Acknowledgements}
This work was supported by European Union under
the RTN contracts HPRN-CT-2000-00148 and
HPRN-CT-2000-00152.

\end{document}